\documentstyle[aps]{revtex}
\begin{document}

\draft

\preprint{gr-qc/9501008}

\title{Bekenstein-Hawking entropy from a phenomenological membrane}

\author{Carsten Gundlach}

\address{LAEFF-INTA, PO Box 50727, 28080 Madrid, Spain}

\date{9 January 1995}

\maketitle

\begin{abstract}
It is pointed out that the entropy of a membrane which is quantized
perturbatively around a background position of fixed radius in a black
hole spacetime is equal to the Bekenstein-Hawking entropy, if 1) the
membrane surface is the horizon surface plus one Planck unit, and 2)
its temperature is the Hawking temperature. (This is a comment on
gr-qc 9411037.)
\end{abstract}

This note is a comment on the recent preprint \cite{Lousto}, where a
Nambu-Goto membrane was proposed as the site of black hole entropy,
but where the resulting quantum theory was subsequently confused with
that of a single point particle on the two-sphere $r={\rm const.}$,
resulting in an entropy of order $\ln(M)$. We now give the correct
calculation.  Consider a massless membrane with Nambu-Goto action
(volume of its worldsheet) in the background spacetime of a
Schwarzschild or Reissner-Nordstr\"om black hole. With $r$ and $t$ the
usual Schwarzschild radial and time coordinates and $\theta$ and
$\phi$ the usual polar coordinates in spherical symmetry, let the
membrane worldsheet be parametrized by
$r(t,\theta,\phi)=R(t)+\eta(t,\eta,\phi)$. The zero mode $R(t)$ is
determined by imposing that $\int\eta\sin\theta d\theta d\phi=0$ for
all $t$. We decompose $\eta(t,\theta,\phi)$ into a spherical harmonics
as $\sum_{l,m}\eta_{lm}(t)Y_{lm}(\theta,\phi)$. Expanding to quadratic
order in the $\eta_{lm}$ we obtain a background action plus a system
of harmonic oscillators $\eta_{lm}$ parametrically coupled to the zero
mode $R$. Now we make the assumption that $R(t)$ is fixed at a
constant value $\bar R$. This is of course inconsistent, as there is
no such background solution other than $\bar R$ equal to the horizon
radius. We quantize the $\eta_{lm}$ around this background value. The
energy levels for each independent harmonic oscillator $\eta_{lm}$ are
given by
\begin{equation}
E_{lm}=\left(N_{lm}+{1\over2}\right)\left[E_0^2(\bar R)+E_1^2(\bar
R)l(l+1)\right]^{1/2},
\end{equation}
where $E_1^2(r)=(r-r_-)(r-r_+)r^4$ for a Reissner-Nordstr\"om black
hole, or $E_1^2(r)=(r-2M)/r^3$ for a Schwarzschild black hole.  The
$N_{lm}$ for each $(l,m)$, where $l=1,2,...$ and $m=-l,...,l$, are the
quantum numbers for the system. In the partition function we sum over
all values $N_{lm}=0,1,2...$ for each $(l,m)$. In calculating the
partition function approximately, we expand for large $l$ and $N$, so
that $E_{lm}\sim N_{lm} E_1(\bar R)$, and approximate a sum over $l$
by an integral. (A similar calculation arises in t'Hooft's ``brick
wall'' model \cite{t'Hooft}.) We obtain for the free energy $F=-T\ln
Z=-2E_1^{-2}(\bar R)T^3$ and hence for the entropy $S=-\partial
F/\partial T$
\begin{equation}
S=6E_1^{-2}(\bar R)T^2
\end{equation}
So far this is only quantum mechanics, and Newton's constant has not
yet appeared. To this perturbative level, the energy levels and hence
the entropy are also independent of the membrane tension.  We now
parametrize $\bar R$ as $ \bar R=2M+\delta/(16\pi M) $ (now in Planck
units $G=1$), which means that its area is $\delta$ Planck units
larger than the horizon area. It has been proposed that $\delta\sim 1$
for a phenomological membrane (``stretched horizon''). We obtain
$E_1=(1/4M^2)\sqrt{\delta/4\pi}$. Assuming that the temperature in our
entropy formula is the Hawking temperature, $T=1/(8\pi M)$, we obtain
for the entropy
\begin{equation}
S=6M^2/(\pi\delta)
\end{equation}
We see that a $\delta$ of order 1 gives the Bekenstein-Hawking
entropy.

\end{document}